\documentstyle[11pt,newpasp,twoside,epsf]{article}
\def\edcomment#1{\iffalse\marginpar{\raggedright\sl#1\/}\else\relax\fi}
\reversemarginpar

\begin{document}
\title{Studying pulsars with the SKA and other new facilities}
 \author{Jon Bell}
\affil{ATNF, CSIRO, PO Box 76 Epping NSW 1710 AUSTRALIA; ~~jbell@atnf.csiro.au}

\begin{abstract}
The Square Kilometre Array (SKA) is a proposed next generation radio
telescope.  Between now and 2005 this project is in a technology development
and prototyping phase, with construction likely to begin in $\sim
2010$. This paper describes what the SKA may be like, its key features, the
motivation for building it and where you can access more details about
it. Its is important to see any new facility in context, so other new
facilities are also discussed. Avenues for future extensibility of the SKA
other telescopes are covered, with some emphasis on multiple beam
systems. Some suggestions for useful pulsar experiments and pulsar searches
strategies are summarised. A conclusion is that the SKA may not be the most
cost effective way to search for pulsars and that a 128 beam receiver system
on an Arecibo like telescope working between 1 and 5 GHz may be a more cost
effective approach.
\end{abstract}

\section{Astronomers Dreams}

To meet the goals outlined in the SKA science case (available on
www.ras.ucalgary.ca/SKA), astronomers have drafted the following
specifications for the SKA:\\

\begin{tabular}{ll}
Frequency coverage           & 0.2 -- 20 GHz \\
Collecting area              & 1,000,000 m$^2$ ($A_{e}/T_{sys} = 2 \times 10^{4}$) \\
Simultaneous frequency bands & $\ge$2\\
Resolution at 1.4 GHz        & 0.1 arcsec \\
Field of view at 1.4 GHz     & $1^0 \times 1^0$ \\
Number of simultaneous beams & 100 \\
\end{tabular}

\vspace{0.5cm}

This is a subset of the more detailed specifications which are available on
the various SKA web pages listed in Section 8. The specifications are being
refined as the science case is refined and it is not too early for people to
be contributing their ideas to the science case.

\section{Potential Construction Methods}

The proposed construction methods may be of some interest to astronomers,
because they lead to considerably different capabilities.  Existing large
radio telescopes like Jodrell Bank, Bonn and Parkes cost thousands of
dollars per square metre to build. To be realistic, the total costof the SKA
has to be no more that 1 Billion US dollars. This implies then that the
collecting area of the telescope has to be built for 200-300 million dollars
or $\sim $\$200 per square metre. This section summarises some of the
methods proposed for achieving this. More details are on the SKA web pages
listed in Section 8.

\begin{description}

\item {\bf KARST } (China) 10--14 dishes of 300--500m in diameter like
Arecibo. The aim being to take advantage of naturally occuring holes in the
ground of suitable shape. Substantially improved sky coverage would be
obtained by only using part of the dish (at any one time) which is
dynamically deformed to focus on a vastly more mobile feed system. Greatly
improved sensitivity by using multiple beam feed/receiver systems. This
method is a good candidate (especially for pulsars) , but the small number
of stations (10--14) does not measure up to the desired image fidelity which
requires 500--1000 stations.

\item {\bf LAR - Large Adaptive Reflector} (Canada) Similar to KARST, but
uses a much shallower parabola, requiring much greater deformations of the
surface and the feed/receiver system to be supported at a height of several
kilometres on a tethered ballon. It gains some sensitivity by using more of
the surface, but suffers from foreshortening at low elevations. It is
unproven technology which is unlikely to be as cheap as the FAST proposal
due to greater engineering complexity and offers no clear advantages over
the FAST proposal apart from more flexibility in location.

\item {\bf Scaled GMRT technology} (India) 9000 12m dishes
built using wire rope and cheap Indian labour. A more conventional design,
which in some respects could be considered the current pace setter as it
satisfies a large fraction of the specifications and the affordability has
already been demonstrated in the construction of GMRT which has 30 45m
dishes and cost $\sim$ \$500 m$^{-2}$. The success or otherwise of GMRT in
the coming years may have a strong bearing on the future success of this
proposal for the SKA.

\item {\bf Commercial 5m Dishes} (USA) 50000 5m commercial off the shelf
dishes. The idea being to take advantage of the cost savings from using
components that are already under mass production. Apart from the
construction method of the dishes, this proposal has a lot in common with
the Indian proposal. The prototype presently being constructed, the 1hT (1
hectare telescope), should give a good indication of the affordability.

\item {\bf Planar Phased Arrays} (Netherlands) This proposal aims to
take the concept of riding on components with commercially driven costs a
step further, by moving as much as possible of the telescope infrastructure
into a technology that not only has high commercial demand, also has a
rapidly improving cost curve and future scalability. ie to move away from
mechanically based infrastructure to information based technology that
presently has an exponential growth curve. The main proposal is to have a
flat collecting area and steer the telescope electronically by forming beams
in the appropriate direction. A great advantage of this proposal is that it
makes it possible to do several experiment simultaniously, by using
multiple beams. However it does suffer from foreshortening, and reduced
sensitivity, because it cannot have cooled receivers.

\item {\bf Luneberg Lenses} (Australia \& Russia) While the above proposals
have been around for some time, this one is quite recent. People are wrapped
with the flexibility and multiple beams of the planar arrays, but frustrated
with the poorer sensitivity and complexity of the planar pahsed
arrays. Luneberg lenses may provide a way to overcome that. A Luneberg lense
is a sphere of dialectirc material which has a radial gradient of refractive
index. A plane wave impinging on the sphere is focussed at a point on the
other side of the sphere - making it possible to have multiple cooled
receivers without enourmous beamforming networks.

\end{description}

\section{Future Growth - Multiple Beams}

To date the sensitivity of radio telescopes has followed an exponential
growth curve. The first and most obvious point about exponential growth is
that it cannot be sustained indefinitely.  Can we maintain it for sometime
into the future ? There are 2 basic ways to stay on this exponential curve:
1) Spend more money which requires bigger and bigger collaborations, 2) Take
advantage of technological advances in other areas. Radio astronomy is at
the point where it needs to do both 1) and 2) to stay on the curve !

\subsection{Extensibility Through Improved Technologies}

\begin{description}

\item{\bf System Temperature:} Reber started out with a 5000 K system
temperature. Modern systems now run at around 20 K, meaning that if
everything else was kept constant, Reber's telescope would now be 250
times more sensitive than when first built. There are possibilities of
some improvements in future, but nothing like what was possible in the
past.

\item{\bf Band Width:} Telescopes like the GBT (Green Bank Telescope)
having bandwidths some 500 times greater than Reber's, will give
factors of 20--25 improvement in sensitivity. Some future improvements
will be possible, but again they will not be as large as in the past.

\item{\bf Multiple Beams:} In the focal or aperture plane, multiple beam
systems provide an excellent extensibility path, allowing vastly deeper
surveys than were possible in the past. Although multiple beam systems have
been used in the past, the full potential of this approach is yet to be
exploited. A notable example that has made a stride forward in this
direction is the Parkes 13 beam L band system (Stavely-Smith et al. 1996,
pASA 13 243). The fully sampled focal plane phased array system being
developed at NRAO by Fisher and Bradley highlights the likely path for the
future.

\end{description}

The sensitivity of the 64m Parkes telescope for example, has improved by a
factor of 400 since 1962. Scope for continuing this evolution looks good for
the next decade, but beyond that more collecting area will be
needed. Putting a 100 beam system on Arecibo by 2005 is technically feasible
and would allow Arecibo to jump out in front of the curve as it did when
first built in 1964.

\section{Other New Facilities}

The aim of this and the following section is to give you a feel for which
new radio facilities are likely to be the most useful for pulsar
studies. Other spectral bands are covered by other speakers at this meeting,
eg Trumper on X-rays. Table 1 summaries the new radio facilities and their
likely completion dates. At present, Arecibo and Parkes (with the 13 beam
receiver system) are roughly equivelent in terms of their survey capability
(rate at which a given area can be searched to a given sensitivity) and are
well ahead of the next best facilities and some of the new facilties such as
ALMA, GBT, 1hT and VLA. GMRT may join this group, as could GBT or any of the
other large single dishes if they opted for multi beam receiver systems.

\begin{table}[h]
\begin{center}
\begin{tabular}{lllll}
      & D(m)            & Area(m$^2$)       & Freq(GHz)   & Date \\
ALMA  & 64 $\times$ 12m & 7.2 $\times 10^3$ & 30.0 -- 900 & 2007\\
GBT   & 1 $\times$ 100m & 7.8 $\times 10^3$ & 0.30 -- 86  & 2000\\
1hT   & 512 $\times$ 5m & 1.0 $\times 10^4$ & 1.00 -- 12  & 2003\\
VLA   & 27 $\times$ 25m & 1.3 $\times 10^4$ & 0.20 -- 50  & 2002\\
GMRT  & 30 $\times$ 45m & 3.0 $\times 10^4$ & 0.03 -- 1.5 & 1999\\
SKA   & ????????        & 1.0 $\times 10^6$ & 0.20 -- 20  & 2015\\
LOFAR & 10$^6$ $\times$ 1m & 1.0 $\times 10^6$ & 0.03 -- 0.2 & 2003\\
\end{tabular}
\caption{New radio astronomy facilities, listed roughly in order of
decreasing frequency and increasing area. The VLA specs include the proposed
major upgrade.}
\label{t:new}
\end{center}
\end{table}

\section{Pulsar Survey Strategies}

The SKA is almost certain to be an array with a large number of elements
(N).  In the past, large area surveys for pulsars at radio wavelengths have
been dominated by large single dishes, Molongolo being a notable
exception. This is mainly because when an array is used to form coherent
beams, the beam size is very small and therefore sky coverage is slow. A
much faster alternative (despite the $sqrt(N)$ loss of sensitivity) is to
form incoherent sums of the N array outputs, thereby surveying the whole
primary beam. However, this requires a detection system (eg a filter bank)
for every antenna element, which becomes expensive. For example (see Table
2) a better rate is achieved in a much more cost effective way by putting 13
receivers and detection systems on Parkes, than putting 27 on the VLA ! A
particularly striking fact shown in Table 2 is that this may continue to be
true even if the SKA is built ! A 100 beam system on Arecibo can find
pulsars as fast as a 7 beam system on the SKA !  For a few million dollars
Arecibo could be kitted out as the search engine to find all the pulsars,
which can then be timed quickly, using the enhanced sensitivity of coherent
beams on the full SKA. Forming coherent SKA beams would also be the method
of choice for small area searches, for example in globular clusters.

To date, pulsar surveys are both sensitivity and dispersion limited,
including the 1.4GHz Parkes multibeam survey. The SKA, LOFAR and a 100 beam
Arecibo will remove the sensitivity limitation. LOFAR is unlikely to find
vast number of millisecond pulsars because it will continue to be dispersion
limited. The SKA and a 100 beam Arecibo can push to GHz frequencies and beat
the dispersion, scattering and sky noise limits while still having enough
sensitivity. This is in contrast to the common (and incorrect) assumption
that pulsars provide a scientific driver for a low frequency SKA.

\begin{table}[h]
\begin{center}
\begin{tabular}{lrrrrr}\hline
Telescope       & Diameter(m)  & Number  & Beams  & Rate  \\ \hline
Present		&	       &         &        &            \\ \hline
ATCA            & 22           & 6       & 1      & 0.2\%      \\
Parkes          & 64           & 1       & 1      & 0.3\%      \\ 
Jodrell         & 76           & 1       & 1      & 0.5\%      \\
WSRT            & 25           & 14      & 1      & 0.7\%      \\ 
Effelsberg      & 100          & 1       & 1      & 0.8\%      \\
VLA             & 25           & 27      & 1      & 1.3\%      \\
Parkes          & 64           & 1       & 13     & 4.2\%      \\ 
Arecibo         & 300          & 1       & 1      & 7.1\%      \\ \hline
Future		&	       &         &        &            \\ \hline
ALMA            & 12           & 64      & 1      & 0.7\%      \\
1hT             & 5            & 512     & 1      & 1.0\%      \\
GMRT            & 45           & 30      & 1      & 4.8\%      \\
Parkes          & 64           & 1       & 64     & 20.7\%     \\ 
Jodrell         & 76           & 1       & 64     & 29.1\%     \\
Effelsberg      & 100          & 1       & 64     & 50.4\%     \\
GBT             & 100          & 1       & 64     & 50.4\%     \\
SKA             & 5            & 50768   & 1      & 100.0\%    \\
Arecibo         & 300          & 1       & 64     & 453.8\%    \\
SKA             & 5            & 50768   & 7      & 700.0\%    \\
Arecibo         & 300          & 1       & 100    & 710.0\%    \\ \hline
\end{tabular}
\caption{Estimated rate at which a given area of sky can be surveyed for
pulsars to a given sensitivity using various exisiting, proposed and
potential radio astronomy facilities. The rates are scaled so that 1 beam on
the SKA is 100\%. The current Parkes multibeam survey (see Camilo et al. from
this meeting) with 13 beams provides a benchmark. This table assumes that
every telescope has equal bandwidth and system temperature. The Parkes
multibeam survey has Tsys $\sim$ 30K and a bandwith of 288 MHz.}
\label{t:surv_rate}
\end{center}
\end{table}

\section{Timing}

Except for a few pulsars, the precision of pulsar timing is presently
limited by sensitivity rather than systematics, uncalibrated instrumental
effects or unmodelled drifts in the pulsar signal. Sensitivity will no
longer be a problem with the SKA and systematics will dominate. Trying to
assess what sort of timing precision is achievable is therefore quite
difficult, because we cannot characterise the systematics yet or know
whether they can be dealt with. Pulsar timing is limited by systematics
around 1 $\mu$s or a bit below. It therefore seems likely that the SKA would
allow a large number (perhaps 100's) of pulsars to be timed to this level of
precision. Using 1 beam it might required something like 10 hours to obtain
suitable data to time 500 pulsars at the 1 $\mu$s level. Using 10 beams that
could be randomly positioned, it would take only 1 hour.

Whether the precision can be pushed well below 1 $\mu$s remains to be
seen. As discussed by Britton at this meeting poor calibration of
polarisation is a dominant systematic effect at present and the invariant
profile method he proposes should to help solve that problem. However, the
SKA is likely to have a planar phased array either in the aperture plane or
the focal plane. The polarisation characteristics of these devices are
substantially different to anything we are currently use for timing
pulsars. It would be advisable for somebody to do some experiments to see
how well we can do pulsar timing with such systems. Another question is the
number of pulses (normally 100's) needed to form a stable profile for
timing. This may limit the speed with which a group of pulsars can be timed
with the SKA.

\section{Some Other Pulsar Uses}

What others things may be possible with the SKA that are not possible
now. Of course there will be many that we cannot yet think of, but it does
not hurt to mention a few of those that we know now.  Simultaneous multi
frequency, multi pulsar studies will be a lot easier. The SKA will make it
possible to do single pulse studies for a large number of pulsars, compared
to the handful at present. One interesting possibility arises if the SKA is
made using a planar phased array or Luneberg lens. One could collect signals
from every element and store them in a FIFO buffer of say 1 hour in
length. In the mean time, collect timing points on pulsars that are likely
to glitch. When one sees that a pulsar has glitched, the FIFO buffer should
be saved so that a beam can then be formed towards the pulsar and thereby
obtain data during the glitch. This idea of course extends beyond pulsars
and could be used for detecting any transient source where there is a
trigger available by other means.

\section{SKA References and Web Resources}
\label{res}

\small{
\begin{center}
\begin{tabular}{lll}\hline
 Reference	& Web Address				& Item of Interest\\\hline
 Australia	&www.atnf.csiro.au/SKA/			&Luneberg Lens\\
 Netherlands	&www.nfra.nl/skai/			&Planar Array\\
 Canada	        &www.ras.ucalgary.ca/SKA/		&Science Case\\
 SETI		&www.seti.org/				&1hT\\
 China		&159.226.63.50/bao/LT			&FAST/KARST
\\
 USA		&www.usska.org				& USA SKA consortium\\
 Backer	        &www.nfra.nl	        &1999 Amsterdam SKA meeting\\
 Bailes	        &www.atnf.csiro.au/SKA/WS/wsmb		&1997 Sydney SKA
 meeting\\
 Fisher	        &www.nrao.edu.au/$\sim$rfisher/		&Array Feed\\\hline
\end{tabular}
\end{center}
}
\end{document}